\documentclass[aps,prd,twocolumn,groupedaddress,showpacs,showkeys]{revtex4-1}

\usepackage{graphicx}

\begin{document}


\title{The Alcubierre Warp Drive: On the Matter of Matter}


\author{Brendan McMonigal}
\email[]{monigal@physics.usyd.edu.au}
\author{Geraint F. Lewis}
\email[]{gfl@physics.usyd.edu.au}
\homepage[]{www.physics.usyd.edu.au/~gfl/}
\author{Philip O'Byrne}
\email[]{poby3060@uni.sydney.edu.au} 
\affiliation{Sydney Institute for Astronomy, School of Physics, A28, The University of Sydney, NSW 2006, Australia}


\date{\today}

\begin{abstract}
The Alcubierre warp drive allows a spaceship to travel at an arbitrarily large global velocity by deforming the spacetime in a bubble around the spaceship.
Little is known about the interactions between massive particles and the Alcubierre warp drive, or the effects of an accelerating or decelerating warp bubble.
We examine geodesics representative of the paths of null and massive particles with a range of initial velocities from \(-c\) to \(c\) interacting with an Alcubierre warp bubble travelling at a range of globally subluminal and superluminal velocities on both constant and variable velocity paths.
The key results for null particles match what would be expected of massive test particles as they approach \( \pm c\). The increase in energy for massive and null particles is calculated in terms of \(v_{s}\), the global ship velocity, and \(v_{p}\), the initial velocity of the particle with respect to the rest frame of the origin/destination of the ship. 
Particles with positive \(v_{p}\) obtain extremely high energy and velocity and become ``time locked" for the duration of their time in the bubble, experiencing very little proper time between entering and eventually leaving the bubble. 
When interacting with an accelerating bubble, any particles within the bubble at the time receive a velocity boost that increases or decreases the magnitude of their velocity if the particle is moving towards the front or rear of the bubble respectively. If the bubble is decelerating, the opposite effect is observed. Thus Eulerian matter is unaffected by bubble accelerations/decelerations. The magnitude of the velocity boosts scales with the magnitude of the bubble acceleration/deceleration.

\end{abstract}


\maketitle


\section{Introduction}
The fundamental limit on the speed of particles implied by Special Relativity has long thought to be a limit to how humans can explore the cosmos.
However, with the deformation of spacetime permitted by General Relativity, globally superluminal movement is possible. 
The Alcubierre Warp Drive spacetime \cite{1} allows a ship to travel between two locations at an arbitrarily large velocity as measured by observers on the ship, as well as at the origin and destination. 

Only a small number of papers have examined the interactions of light with Alcubierre warp bubbles \citep{17, c33, c27, c21, c22}, with detailed analysis only by \citet{8}.
Furthermore there has been almost no coverage in the literature of the interactions of massive particles with warp bubbles. \citet{4} investigate these interactions only briefly, discussing only the interaction between a warp bubble travelling at constant velocity and Eulerian matter, that is matter stationary in the rest frame of the origin/destination of the ship.  This paper fills this void, providing a detailed analysis of the interactions of null and massive particles with an Alcubierre warp bubble at both constant and variable velocity, via the calculation of representative geodesics through Alcubierre spacetime.

The outline of the paper is as follows. In section \ref{background} we introduce the Alcubierre spacetime \cite{1} and discuss the main concerns regarding its validity that have been proposed. In section \ref{main} we outline the variable velocity paths used and equations of motion before presenting our analysis for interactions with bubbles at constant velocity in \ref{section_constant}. We extend this to one way trips in section \ref{section_oneway} and rounds trips in \ref{section_round}. In section \ref{end} we summarize and conclude.

\section{Alcubierre Warp Drive}\label{background}

The Alcubierre spacetime is asymptotically flat with the exception of the walls of a small spherical bubble surrounding supposedly a ship. The general idea behind the Alcubierre warp drive is to choose an arbitrary path and deform spacetime in the immediate vicinity such that the path becomes a timelike freefall path i.e. a geodesic. Since the ship is following a geodesic, the travellers experience no inertial effects.

The Alcubierre metric \cite{1} can be described by 
\begin{equation}
ds^{2} = -dt^{2} + (dx - v_{s}(t)f(r_{s})dt)^{2} + dy^{2} + dz^{2},
\end{equation}
where \( f \) is the shape function which determines the form of the bubble wall spacetime distortion, normalised to unit value at the centre of the bubble. In Alcubierre's original paper it was proposed
\begin{equation}
f(r_{s}) = \frac{tanh(\sigma(r_{s} + R)) - tanh(\sigma(r_{s} - R))}{2tanh(\sigma R)},
\end{equation}
where \( r_{s}(t) = \sqrt{(x - x_{s}(t))^{2} + y^{2} + z^{2}} \) is the distance from the ship, \( v_{s}(t) \) is the global velocity of the ship, and hence the bubble, and can be arbitrarily large, and R and \( \sigma \) are arbitrary parameters determining the radius of the warp bubble and the thickness of the bubble wall respectively.
For the purposes of this paper, a qualitatively similar but mathematically more manageable equation was used, namely \( f(r_{s}) = 1 - \left( \frac{r_{s}}{R} \right) ^{4} \) for \(r_{s}<R\) and 0 otherwise.

The global velocity is that as measured by Eulerian observers, that is anyone in flat spacetime and in the rest frame of the origin/destination of the ship. It is straightforward to see from the line element that the form of global velocity is \( v(t) = \frac{dx(t)}{dt} \). From this definition of global velocity, it is also straightforward to see from the line element, that \( d\tau = dt \) for the travellers on the ship, where \(\tau\) is proper time. Thus the proper time of the Eulerian observers and the travellers is the same, and hence no time dilation is experienced. 

Alcubierre produced this solution by what is termed ``metric engineering'', that is, stipulating the required spacetime geometry and solving for the necessary energy distribution. This method is problematic as it can result in seemingly unphysical solutions.
As noted by Alcubierre, the stress energy momentum tensor is negative for all observers, even when operating at arbitrarily low velocity \citep{14, 22}. This implies a requirement of negative energy densities,
which can only be achieved by exotic matter\cite{[{A detailed discussion of exotic matter and the validity of warp drive spacetimes in general is given by }]c29} thus violating the Energy Conditions \citep{s11}. 
This violation implies ability to create Closed Timelike Curves (CTCs) and their associated problems. In addition to these issues, the Alcubierre Warp Drive suffers from the Tachyonic Problem as described by \citet{5}.

To circumvent these problems, modifications the metric have been proposed, however any spacetime that permits apparent superluminal travel violates the Weak and Null energy conditions, and by extension opens the way for CTCs and their associated problems \citep{s8, s9, s10}. 
 \citet{17} suggest that even if these problems were viewed simply as engineering issues, there would still be critical problems due to semiclassical instability.


\section{The Influence on Particles} \label{main}

The following analysis is restricted to the t-x plane, and thus \(r_{s}(t) \) simplifies to the signed distance from the ship \(x - x_{s}(t)\). First we look at the interactions of null and massive particles with a bubble at constant velocity, then with a bubble on a one way trip, and finally with a bubble on a round trip.

\subsubsection{Non-uniform Paths}

Two types of variable velocity path are used in this paper. One way trips characterised by a logistic curve of the form in equation \ref{logisticx} and velocity given by equation \ref{logisticv}, and round trips characterised by gaussian functions of the form in equation \ref{gaussianx} with velocity given by equation \ref{gaussianv}.

\begin{equation}
x_{s}(t) = \frac{b}{1+exp(\frac{d-t}{a})}+e\label{logisticx}
\end{equation}
\begin{equation}
v_{s}(t) = \frac{dx_{s}(t)}{dt} = \frac{b \, exp(\frac{d-t}{a})}{a(1+exp(\frac{d-t}{a}))^{2}}\label{logisticv}
\end{equation}
%
\begin{equation}
x_{s}(t) = b \, exp\left(-\left(\frac{t-d}{a}\right)^{s}\right)+e\label{gaussianx}
\end{equation}
\begin{equation}
v_{s}(t) = \frac{dx_{s}(t)}{dt} = -\frac{sb}{a}exp\left(-\left(\frac{t-d}{a}\right)^{s}\right)\left(\frac{t-d}{a}\right)^{s-1}\label{gaussianv}
\end{equation}
In each path, the parameters a,b,d, and e respectively set the path length in t, the path length in x, the midpoint in t, and the journey origin in x. For the gaussian path, there is an additional parameter s, which is an even positive integer. For larger s values, the path more closely resembles a top hat.



\subsubsection{Equations of Motion}
The geodesic equations
\begin{equation}
\frac{d^{2}x^{\alpha}}{d\lambda^{2}} + \Gamma^{\alpha}_{\beta\gamma} \frac{dx^{\beta}}{d\lambda}\frac{dx^{\gamma}}{d\lambda} = 0,
\end{equation}
describe freefall paths through spacetime. Timelike geodesics are parameterised by \( \tau \) whereas null geodesics must be parameterised by an affine parameter \( \lambda \). The nonzero Christoffel symbols are
\footnote{These equations were numerically integrated to find \(t\), \(x\), \(u^{t}\) and \(u^{x}\) of the particles using the inbuilt Ordinary Differential Equation (ODE) solver ode23s in Matlab. Only one pair of these variables (\(t\) and \(u^{t}\), or \(x\) and \(u^{x}\)) needed to be solved for, however both were found and the normalisation of the 4 velocity was used as a check on the accuracy of the integration. In cases where \(u^{t}\) and \(u^{x}\) diverge, the equations were numerically integrated to find x and \(\frac{dx}{dt}\) using ode45.}


\begin{eqnarray}
\Gamma^{t}_{tt} &=& -\Gamma^{x}_{tx} = -\Gamma^{x}_{xt} = v_{s}^{3}(t)f^{2}(r_{s})\frac{\partial{f(r_{s})}}{\partial{x}}, \label{Christoffel1} \\
\Gamma^{x}_{xx} &=& -\Gamma^{t}_{tx} = -\Gamma^{t}_{xt} = v_{s}^{2}(t)f(r_{s})\frac{\partial{f(r_{s})}}{\partial{x}},  \\
\Gamma^{t}_{xx} &=& v_{s}(t)\frac{\partial{f(r_{s})}}{\partial{x}}, \\
\Gamma^{x}_{tt} &=& v_{s}^{2}(t)f(r_{s})(v_{s}^{2}(t)f^{2}(r_{s})-1)\frac{\partial{f(r_{s})}}{\partial{x}} \label{Christoffel2}\\&~& - f(r_{s})\frac{\partial{v_{s}(t)}}{\partial{t}} - v_{s}(t)\frac{\partial{f(r_{s})}}{\partial{t}} \nonumber
\end{eqnarray}

\subsubsection{Particle Energy}

The energy of the particles at the ship and outside the bubble is calculated using 
\begin{equation}
E = -\mathbf{p} \cdot \mathbf{u_{obs}}\label{energy}
\end{equation}
where \( {u_{obs}^{\alpha}}\) is the 4-velocity of the observer, and \({p^{\alpha}}\) is the 4-momentum of the particle. For an Eulerian observer at rest with respect to the \((t,x)\) coordinate system, the 4-velocity is \( (1,0,0,0) \); interestingly we can show that for the traveller at the centre of the bubble the 4-velocity is also \( (1,0,0,0) \)\cite{energy,4}.
The redshift is then calculated using \(z+1 = \frac{E_{emitted}}{E_{observed}}\). Thus the relative increase or decrease in particle energy is represented by the quantity 
\begin{equation}
b = \frac{E_{observed}}{E_{emitted}} = \frac{1}{z+1}
\end{equation}
 which we define the blueshift.
As this is calculated from the time component of the 4-velocity, it is also representative of the time dilation for massive particles when measured at by Eulerian observers or observers on the ship.

\subsubsection{Horizons}


A useful question to ask is when the global velocity of the particle is larger than the global velocity of the ship.
For null particles, we start with \(ds^{2} = 0\) and rearrange to find \(\frac{dx}{dt} = \pm 1 + v_{s}(t)f(r_{s})\). Substituting into \(v_{p} \ge v_{s}(t)\) and solving for our shape function gives
\begin{equation}
|x_{p}(t) - x_{s}(t)| \le \frac{R}{v_{s}^{\frac{1}{4}}(t)} \label{horizon}
\end{equation}
which defines two positions. We define the position on the side towards which the bubble is moving to be the front horizon and the other to be the rear horizon. Notably, these positions are outside the bubble radius when \(v_{s}(t) < 1\) and thus the horizons only exist while the ship is moving at superluminal velocity.

For massive particles, this question is more difficult.
Starting from the normalisation \(\mathbf{u} \cdot \mathbf{u} = -1 \), we find 
\( (v_{s}^{2}(t)f^{2}(r_{s})-1)u^{t2} - 2v_{s}(t)f(r_{s})u^{t}u^{x} + u^{x2} = -1 \), which using \( \frac{u^{x}}{u^{t}} = \frac{dx}{dt} \) becomes
\(
v_{s}^{2}(t)f^{2}(r_{s})-1 - 2v_{s}(t)f(r_{s})v_{p} + v_{p}^{2} = -\frac{1}{(u^{t})^{2}}\label{pre-f}
\)
We found that when a bubble catches up to a particle with a nonzero global velocity in the same direction as the ship, \(u^{t}\) diverges, which implies \( -\frac{1}{(u^{t})^{2}} \to 0\). 
This returns us to the same equation as for null particles, and thus the same equation for the horizons, but due to our assumption that the particle is interacting with the bubble from the front, this can only tell us about the front horizon. 
We revisit the rear horizon for massive particles later.

\subsection{Constant Velocity Bubbles}\label{section_constant}

\subsubsection{Null Particles}
 
\begin{figure}[tbp]
\includegraphics[width=79mm]{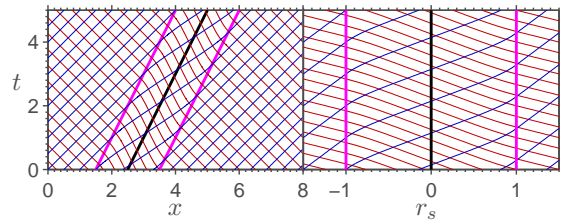}
\caption[Light interacting with a subluminal bubble]{The left panel is a spacetime diagram of a ship which has been travelling at a constant subluminal velocity of 0.5 interacting with light rays from outside the bubble. The abscissa represents the \(x\) coordinate and the ordinate represents the \(t\) coordinate. The aspect ratio is 1:1 such that light rays travel at \(45^{\circ}\) in flat spacetime. The right panel is the same interaction with respect to the ship. The abscissa represents the distance with respect to the ship, and the ordinate represents the \(t\) coordinate. The blue and red lines are forward and backward travelling light rays respectively, and the magenta and black lines represent the bubble walls and ship respectively. }
\label{linear light sub}
\end{figure}

\begin{figure}[tbp]
\includegraphics[width=79mm]{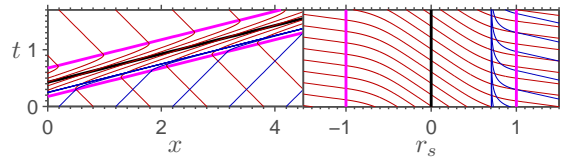}
\caption[Light interacting with a superluminal bubble]{As in Figure \ref{linear light sub} for a constant superluminal bubble velocity of 4. \label{linear light sup}}
\end{figure}

Figure \ref{linear light sub} shows light rays interacting with a warp bubble travelling at a constant subluminal global velocity of 0.5. When the bubble velocity is subluminal, all forward and backward moving light rays pass through the bubble, however the forward moving light rays take longer to do so with respect to both observers on the ship and Eulerian observers outside the bubble. The light rays exit the bubble with the same frequency with which they enter it, however they are spatially displaced by a small distance in the direction of travel of the ship. The magnitude of the displacement is due to the time spent in the bubble and the bubble velocity, larger for longer time spent in the bubble and for larger ship velocity. Thus the displacement for forward travelling light rays is slightly larger than that of backward travelling light rays. 

Figure \ref{linear light sup} shows the same situation for a warp bubble of constant superluminal velocity 4. The backward travelling light rays still pass through the bubble, exiting the bubble with the same frequency with which they enter it. As the bubble catches up to the forward travelling light rays, they enter the bubble and asymptote towards the position given by Equation \ref{horizon} corresponding to the front horizon before reaching the ship. Due to this, the space behind the bubble is virtually devoid of forward travelling light rays.

\subsubsection{Massive Particles}

\begin{figure}[tbp]
\includegraphics[width=79mm]{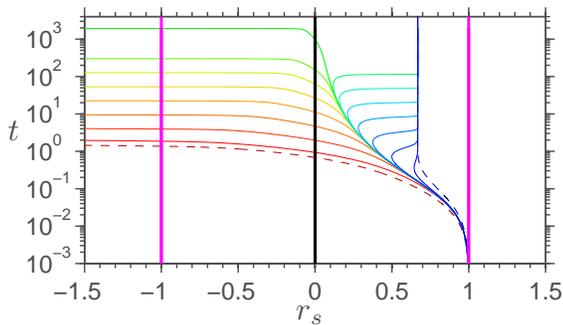} 
\caption[Paths through a superluminal bubble]{Paths of particles interacting with a bubble of constant superluminal velocity 5. The abscissa is the distance with respect to the ship, and the ordinate is the \(t\) coordinate. All particles enter the bubble at \(t=0\). Going clockwise from the bottom left, the paths are for backward travelling light, \(v_{p} = -10^{-1} ~\text{to}~ -10^{-7}\) in unit powers of \(10\), \(v_{p}=0,~ v_{p} = 10^{-7} ~\text{to}~ 10^{-1}\) in unit powers of \(10\), and forward travelling light. The bubble and ship are marked as in Figure \ref{linear light sub}.
 \label{linear matter bubble sup}}
\end{figure}


Figure \ref{linear matter bubble sup} shows the paths taken by massive particles with respect to the ship when interacting with a bubble of superluminal velocity 5. When the initial particle velocity is zero, the particle passes through the bubble, which agrees with the result presented by \citet{4}. This is the path which takes the longest coordinate time to reach the ship and subsequently leave the rear of the bubble. Of the paths that leave the bubble, no other path spends more coordinate time in the bubble and hence this path results in maximal spatial displacement. Particles with negative initial velocity have a path between that of zero velocity particles and backward travelling light. The larger the magnitude of the negative velocity, the earlier the path diverges from that taken by zero velocity particles and the more closely the path approximates that taken by backward travelling light. As the magnitude of the negative velocity decreases, the path taken diverges from that taken by zero velocity particles later, but even particles with a negative velocity of only \(-10^{-7}\) spend less coordinate time in the bubble than zero velocity particles by almost a magnitude.

Positive velocity particles similarly have a path between that of zero velocity particles and forward travelling light. Positive velocity particles never reach the ship, and all asymptote to the same position given by Equation \ref{horizon} corresponding to the front horizon. As the magnitude of the velocity increases, the path taken by the particles approximates that taken by forward travelling light, and as the magnitude of the velocity decreases, the path taken follows the zero velocity particle path for longer before rapidly diverging and asymptoting towards the path taken by light.

\begin{figure}[tbp]
\includegraphics[width=79mm]{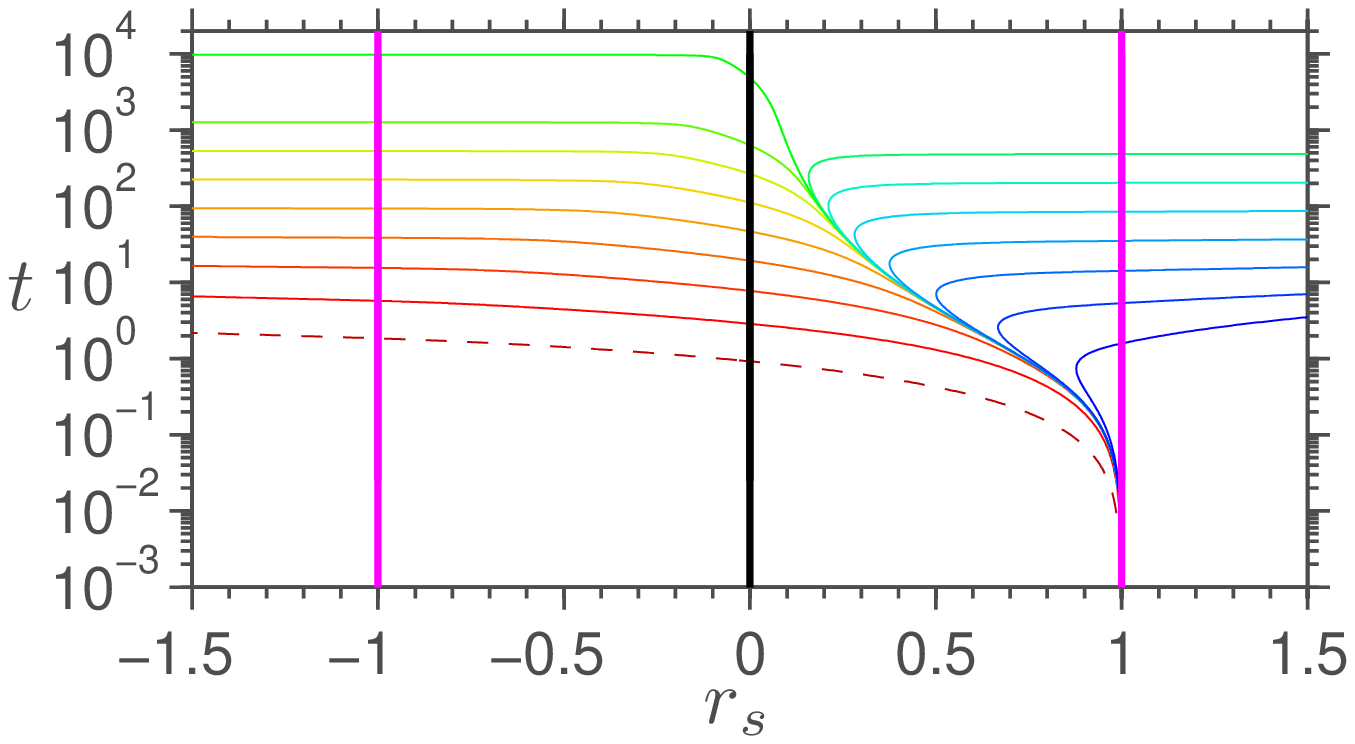}
\includegraphics[width=79mm]{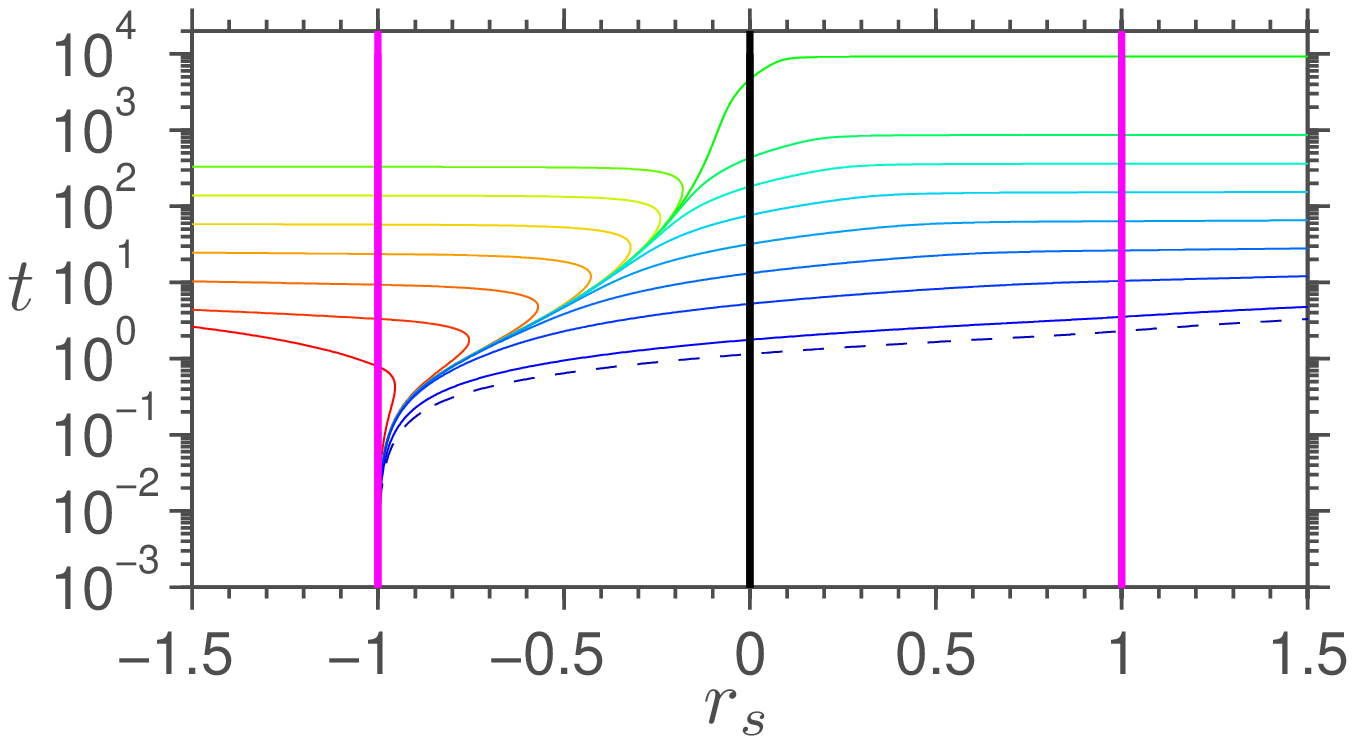}
\caption[Paths through a subluminal bubble]{Paths of particles interacting with a bubble constant subluminal velocity 0.5. The abscissa, ordinate, ship and bubble are as in Figure \ref{linear matter bubble sup}. In the upper panel, the paths are as in Figure \ref{linear matter bubble sup} with the exclusion of forward travelling light as it does not interact with a subluminal bubble from the front. In the lower panel, the paths are for velocities larger than those of Figure \ref{linear matter bubble sup} by the magnitude of the critical velocity (which in this case is 0.8) with the exclusion of backward travelling light as it does not interact with a subluminal bubble from behind; they are distributed the same way.}
\label{linear matter bubble sub}
\end{figure}

\begin{figure}[tbp]
\includegraphics[width=79mm]{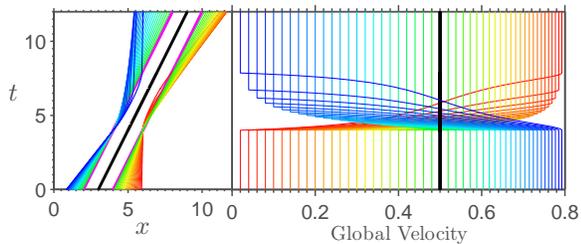}
\caption[``Slow'' matter interactions]{``Slow'' matter interactions with a subluminal bubble of velocity 0.5. The left panel is a spacetime diagram as in Figure \ref{linear light sub} where the ordinate represents the \(t\) coordinate and the abscissa represents the \(x\) coordinate. The right panel shows the same interaction where the abscissa represents the global velocity. The interactions shown are for particles with velocities from \(v_{p} = 0.02 ~\text{to}~ 0.48\) in steps of \(0.02\) and their corresponding final velocities.}
\label{linear matter slow}
\end{figure}

Figure \ref{linear matter bubble sub} shows the paths taken by massive particles with respect to the ship when interacting with a bubble of subluminal velocity 0.5. Particles with non positive initial velocity pass through the bubble as for superluminal bubble interactions, but take longer to do so with the zero velocity path taking almost a magnitude longer than for the 5c bubble interaction. Positive initial velocity particle paths are divided into paths which resemble that taken by forward travelling light and paths which we refer to as ``slow'' matter paths; this division occurs at a velocity we define the critical velocity, \(v_{c}\). Particles travelling at the critical velocity are the slowest particles which pass through the bubble from behind. This is similar to how zero velocity particles represent the slowest path which passes through the bubble from infront. These particles are the slowest in the sense that their velocity magnitude is the lowest, and they take the largest amount of coordinate time to pass through the bubble. As the velocity increases above the critical velocity, the path taken diverges from that of particles at the critical velocity earlier, more closely approximating the path taken by forward travelling light.

For particles with initial positive velocity lower than the critical velocity, there are two possibilities, either the velocity is above that of the bubble and hence the particle will catch up to the bubble from behind or the velocity is below that of the bubble and thus the particle will interact with the front of the bubble first as the bubble catches up to the particle. 
In the first case, the particle is ejected from the bubble back out the rear with a reduced but still positive velocity below the ship velocity. For particles with initial velocity closer to the critical velocity, the path follows that of the critical velocity particles for longer, thus getting closer to the ship, before diverging and being ejected from the bubble. 
Similarly in the second case, the particle is ejected out the front of the bubble with an increased velocity larger than the ship velocity. For particles with a smaller initial velocity, the path follows that of the zero velocity particles for longer, thus getting closer to the ship, before diverging and being ejected from the bubble.


While the particular results presented are for our choice of shape function only, we can generalise to any shape function consisting of expansion behind the ship and contraction infront of the ship. We found that particles truly at rest before entering the warp bubble come to rest when they reach a region of flat spacetime again; the results presented suggest that this first occurs as the particle leaves the bubble, but infact it occurs upon reaching the flat spacetime region immediately surrounding the ship. Our choice of shape function restricts this region to a single point and thus  our use of numerical methods does not allow it to land directly on this point, instead the particle passes the ship as if it had a infinitesimal negative velocity. We found that while within regions of contracting space, the magnitude of the velocity of particles increases. This means that particles with negative velocity accelerate through the front half of the bubble, and that particles with positive velocity accelerate forwards away from the ship. Thus any shape function of this form will result in particles with zero initial velocity coming to rest at the ship, separating the paths which pass through the bubble from those which do not.

Figure \ref{linear matter slow} shows ``slow'' matter paths for a bubble velocity of 0.5. The closer the initial velocity is to the bubble velocity, the closer the final velocity is to the bubble velocity. Furthermore, there is a symmetry between the two interactions such that if a particle has initial velocity \(0 < v_{p1} < v_{s}\), then the bubble will catch up to the particle and eject it with a new velocity \(v_{s} < v_{p2} < v_{c}\). If this particle were to catch up to another warp drive travelling at the same velocity (and thus having the same critical velocity), then after entering the bubble it would be ejected out the rear with final velocity \(v_{p1}\); this symmetry is evident in Figure \ref{linear matter slow}. This symmetry is not evident in Figure \ref{linear matter bubble sub} as the final velocities obtained by the ``slow'' matter in Figure \ref{linear matter bubble sub} interacting from the front do not correspond to the initial velocities of the ``slow'' matter in Figure \ref{linear matter bubble sub} interacting from behind, and vice versa. 

\subsubsection{Blueshifts}

\begin{figure}[t]
\includegraphics[width=79mm]{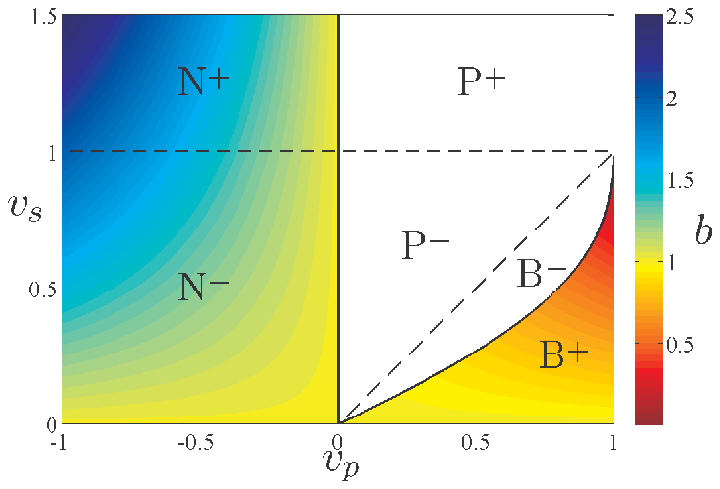}
\includegraphics[width=79mm]{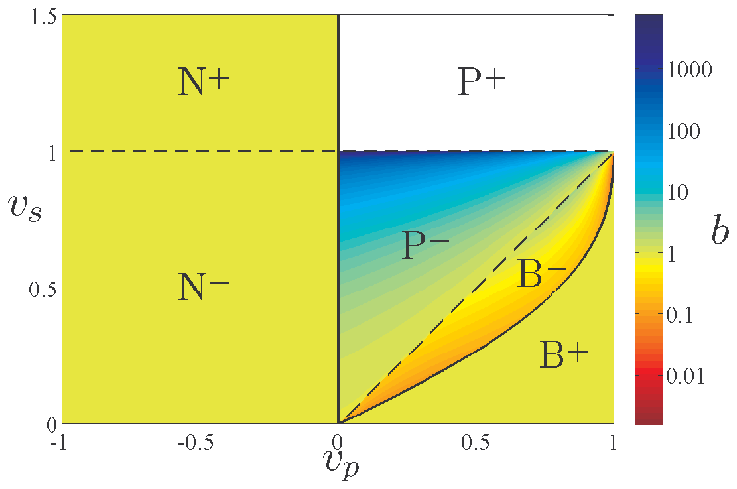}
\caption[Blueshifts observed at the ship and outside the bubble]{(Image quality has been reduced for arXiv submission) The upper panel is blueshifts seen by observers on the ship. The lower panel is for blueshifts seen outside the bubble by observers stationary with respect to the origin. \label{vsvd both}}
\end{figure}

The different interactions that can occur between particles and a warp bubble are summarized in Figure \ref{vsvd both}. N\(^{+}\) and N\(^{-}\)represent particles with initial negative velocity meeting a superluminal and subluminal bubble respectively, and passing through it. P\(^{+}\) and P\(^{-}\) represent particles with initial positive velocity which are overtaken by a superluminal and subluminal bubble respectively, and are subsequently captured in the front of the bubble and ejected from the front of the bubble respectively. B\(^{+}\) and B\(^{-}\) represent particles with initial positive velocity greater than the ship velocity, which catch up with a subluminal bubble from behind, and pass through the bubble and are released out the back of the bubble with a reduced positive velocity respectively. The curved line separating B\(^{+}\) and B\(^{-}\) marks the critical velocity of the bubble. 
It is important to remember that while the ship velocity is only depicted up to \(v_{s} = 1.5\), it can be arbitrarily large. No interactions occur along the line \(v_{p} = v_{s}\) as the particles and the bubble never meet. The line \(v_{p} = 0\) is included only in the regions N\(^{+}\) and N\(^{-}\), and similarly the line \(v_{p} = v_{c}\) is included only in the region B\(^{+}\). Thus P\(^{-}\) and B\(^{-}\) are open regions. The lines \(v_{p} = -1\) and \(v_{p} = 1\) correspond to backward and forward travelling null particles respectively and can be thought of as being part of the regions they bound. Thus N\(^{-}\) and B\(^{+}\) are closed regions. The regions that contain null paths, i.e. N\(^{+}\), P\(^{+}\), N\(^{-}\) and B\(^{+}\), represent all light-like paths. The regions which do not contain null paths, i.e. P\(^{-}\) and B\(^{-}\) represent all ``slow'' matter paths.

The following blueshift results apply to both null and massive particles.
The top image in Figure \ref{vsvd both} shows the blueshift observed at the ship for all particle interactions. Particles in the P\(^{+}\) are captured in the front of the bubble and never reach the ship, and similarly particles in the ``slow'' matter regions are ejected from the bubble without ever reaching the ship, thus there is no value for these regions. Particles in the B\(^{+}\) region, i.e. catching up to the ship from behind, have a blueshift of \(0 < b \le 1\) which is unity only when the ship velocity is 0. As the ship velocity increases towards \(1\), the observed blueshift decreases towards the limiting value of 0. The blueshift also increases with the particle velocity but the dependence on the ship velocity is much larger due to the restrictions on the particle velocity for reaching the ship.
Particles in the N regions, i.e. meeting the ship head on, have a blueshift of \(b \ge 1\) which unity when the ship velocity is 0. In addition, the blueshift is also unity whenever the particle velocity is 0. When both the ship velocity and particle velocity are nonzero, increasing the particle velocity towards \(-1\) increases the observed blueshift, and similarly increasing the ship velocity increases the observed blueshift. As there is no bound on the ship velocity in the N\(^{+}\) region, the blueshift observed can be arbitrarily large.
All of the above mentioned blueshifts are given by 
\begin{equation}
b = 1-v_{s}v_{p} \label{blueshift ship}
\end{equation}

The bottom image in Figure \ref{vsvd both} shows the blueshift observed at the origin or destination, or indeed anywhere outside the bubble for an Eulerian observer. The particles which reach the ship pass through the bubble and exit with the same velocity and energy with which they entered, i.e. the blueshift for the regions N\(^{+}\), N\(^{-}\) and B\(^{+}\) is unity.
The blueshift for the ``slow'' matter is more complicated. In the P\(^{-}\), i.e. when a subluminal bubble catches up to particles, the blueshift increases without bound as the ship velocity approaches unity. As the particle velocity increases towards the ship velocity, the blueshift drops towards unity. The blueshifts for the P\(^{-}\) region are given by 
\begin{equation}
b =  1 + \frac{v_{s}-v_{p}}{v_{s}}\left( \gamma_{c} - 1 \right) ~\text{where}~  \gamma_{i} = \frac{1}{\sqrt{1-v^{2}_{i}}}\label{blueshift p-}
\end{equation}

In the B\(^{-}\) region, i.e. when ``slow'' matter catches up to a subluminal bubble, there are two bounds to the blueshift. As the particle velocity approaches the ship velocity from above, the blueshift rises towards unity, whereas as the the particle velocity increases toward the critical velocity, the blueshift drops toward 0. The blueshifts for the B\(^{-}\) region are given by 
\begin{equation}
b =  \frac{\left(1-\gamma_{c}\right)^{2}+\gamma^{2}_{p}v^{2}_{s}}{\gamma_{c}\gamma^{2}_{p}v^{2}_{s} + \gamma_{p}\sqrt{\left(1-\gamma_{c}\right)^{2} + v^{2}_{s}\left(\gamma^{2}_{p}-\gamma^{2}_{c}\right)}}\label{blueshift b-}
\end{equation}

Technically the particles in the P\(^{+}\) cannot leave the bubble, and so there is no value given for this region. However, the time component of the 4-velocity the particles in this region increases exponentially for the duration of the time they are caught in the bubble. Hence if the ship were to ever slow to below the speed of light such that they could escape and interact with outside observers, they would be observed to have extremely large energies.


\subsection{Superluminal One Way Trips}\label{section_oneway}
\subsubsection{Null Particles}\label{one way trip light} 


\begin{figure}[tbp]
\includegraphics[width=79mm]{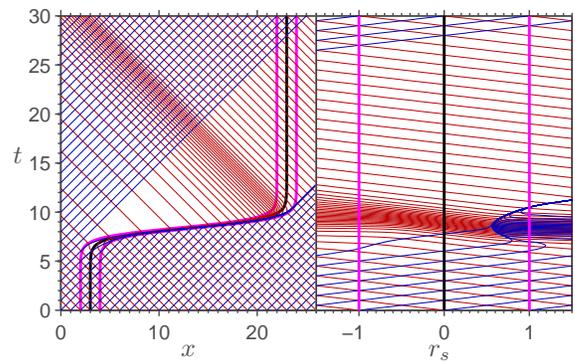}
\caption[Light interacting with a bubble on a superluminal one way trip]{The left panel is a spacetime diagram of light interacting with a ship travelling on a superluminal one way journey. The abscissa represents the \(x\) coordinate and the ordinate represents the \(t\) coordinate. The aspect ratio is 1:1 such that light rays travel at \(45^{\circ}\) in flat spacetime. The blue and red lines are forward and backward travelling light rays respectively, and the magenta and black lines represent the bubble walls and ship respectively. The right panel is the same scenario with respect to the ship, thus the abscissa represents the distance with respect to the ship. }
\label{logistic light sup}
\end{figure}


Figure \ref{logistic light sup} shows 
light interacting with a bubble which reaches a superluminal maximum velocity of 10. 
As the bubble accelerates, the backward travelling light rays are released from the bubble with a spatial separation lower than that they had before entering the bubble. Similarly during deceleration the backward travelling light rays are released from the bubble with a spatial separation lower than their initial spatial separation. These increases and decreases in spatial separation scale with the magnitude of the bubble acceleration and deceleration respectively.

As first glance these effects imply that the spatial separation of the light rays should be unchanged around the centre of the journey, however this does not occur in Figure \ref{logistic light sup} until around 3/4 of the way through the journey. The reason for this is that light rays take time to pass through the bubble, and thus would only appear to be unaffected by acceleration/deceleration if the magnitude of the effects of the acceleration and deceleration were equal. As the period of acceleration (i.e. the first half of the journey) is symmetric to the period of deceleration (i.e. the second half of the journey) on the one way trips in this chapter, the light rays which appear unaffected are those which reach the centre of the bubble at the midpoint of the journey. This occurs for light rays which enter the bubble approximately 1/4 of the way through the journey and exit the bubble approximately 3/4 of the way through the journey, as is observed.

The effects on the change in spatial separation can be explained in terms of the difference in the velocity of the bubble while the light rays are entering and leaving the bubble. Figure \ref{logistic light sup} shows that the rate which the backward travelling light rays enter the bubble is the same as the rate with which they leave the bubble. Thus the change in spatial separation is due only to the difference between the distance that is covered by the ship between picking up light rays and the distance covered between dropping off the light rays.

Figure \ref{logistic light sup} shows that as the bubble catches up to forward travelling light rays, they are captured and asymptote to the front horizon given by Equation \ref{horizon}. The dependence on the ship velocity of the position of this horizon in the bubble is now visible. 
Forward travelling light inside the bubble prior to and during the period of bubble acceleration either asymptotes toward the front horizon in the bubble or is dropped out the rear of the bubble during the journey. If the light is behind the rear horizon, given by Equation \ref{horizon}, then the light will move towards the rear of the bubble and be dropped off upon reaching the bubble edge. If the light is infront of the rear horizon, then it will asymptote towards the front horizon. As the position of the horizons is dependant on the bubble velocity, during periods of large acceleration, forward travelling light within the bubble can be overtaken by the rear horizon. As explained earlier, the horizons only exist for superluminal bubble velocities, and thus upon the bubble velocity decreasing back to subluminal velocity, all the forward travelling light that was previously captured in the front horizon is released at once.

Due to the capture of forward travelling light, superluminally travelling bubbles create a region behind themselves where all forward travelling light from that space has been swept up into the forward horizon. However, due to the release of forward travelling light from behind the rear horizon, this region of space is not entirely devoid of forward travelling light, but contains a sparse distribution of highly redshifted forward travelling light. The range of blueshifts observed at the ship are \( 1 \le b < 10.9 \) for backward travelling light. The range of blueshifts observed outside the bubble are \( 0.257 < b < 3.89 \) for backward travelling light. Blueshifts are not observed for forward travelling light as it does not reach the ship, and does not leave the bubble.
\subsubsection{Massive particle interactions for a superluminal one way trip}\label{matter superluminal one way section}


\begin{figure}[tbp]
\includegraphics[width=79mm]{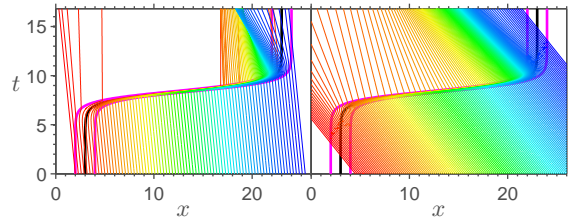}
\caption[Massive particles with small negative velocity interacting with a bubble on a subluminal one way trip]{The left panel is a spacetime diagram of massive particles with initial velocity -0.1 interacting with a ship travelling on a superluminal one way journey. The right panel is the same as the left panel for massive particles with initial velocity -0.8. In each panel, the abscissa represents the \(x\) coordinate and the ordinate represents the \(t\) coordinate. The aspect ratio is 1:1 such that light rays travel at \(45^{\circ}\) in flat spacetime.  The magenta and black lines represent the bubble walls and ship respectively.   \label{logistic matter sup}}
\end{figure}


Figure \ref{logistic matter sup} shows massive particles with negative initial velocity interacting with a warp bubble on a one way trip which reaches a maximum velocity of 10. All the particles pass through the bubble. The rarefaction of the massive particles created during the bubble acceleration and contraction during the deceleration match that shown for the superluminal bubble interacting with light rays in Figure \ref{logistic light sup}, however the change in the energy of the massive particles is now visually apparent via the change in the velocity which can be seen as the gradient when outside the bubble in Figure \ref{logistic matter sup}. Thus we can see that the massive particles which are sparsely released behind the bubble during the acceleration have a reduced velocity and the massive particles which are released rapidly during deceleration have an increased velocity. 
%
The magnitude of the velocity increase and blueshift obtained scales with the magnitude of the intial velocity. The final velocity and blueshifts obtained increase towards those of backward travelling light as the initial velocity approaches \(-1\).


\begin{figure}[tbp]
\includegraphics[width=79mm]{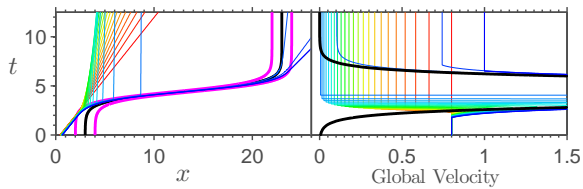}
\caption[Massive particles with positive velocity interacting with a bubble accelerating to superluminal velocity]{The left panel is a spacetime diagram of a ship on a superluminal one way trip interacting with massive particles with initial velocity 0.8 catching up to the bubble as the ship accelerates away to superluminal velocity. The abscissa represents the \(x\) coordinate and the ordinate represents the \(t\) coordinate. The aspect ratio is 1:1 such that light rays travel at \(45^{\circ}\) in flat spacetime. The right panel is same interaction showing the global velocity of each particle and the ship, thus the abscissa represents global velocity. The magenta and black lines represent the bubble walls and ship respectively. }
\label{logistic matter sup rear}
\end{figure}

\begin{figure}[tbp]
\includegraphics[width=79mm]{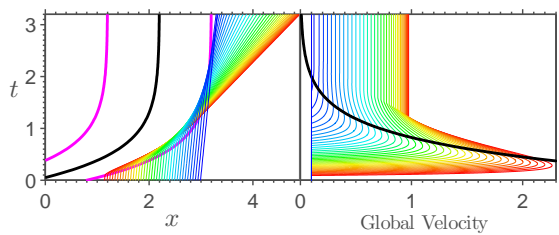}

\caption[Massive particles with positive velocity interacting with a bubble decelerating from superluminal velocity]{The left panel is a spacetime diagram of a ship on a superluminal one way trip interacting with massive particles with initial velocity 0.1. The bubble catches up to the massive particles as it decelerates from superluminal velocity. The abscissa represents the \(x\) coordinate and the ordinate represents the \(t\) coordinate. The aspect ratio is 1:1 such that light rays travel at \(45^{\circ}\) in flat spacetime. The right panel is the same interaction showing the global velocity of each particle and the ship, thus the abscissa represents global velocity. The magenta and black lines represent the bubble walls and ship respectively. }
\label{logistic matter sup front}
\end{figure}


As discussed earlier, all massive particles with positive initial velocity interacting with a superluminal bubble from the front asymptote to the front horizon and receive an exponential increase in energy. However, as was mentioned with light, forward travelling massive particles which are already within the bubble during the acceleration to superluminal velocities either continue forwards and asymptote to the front horizon or move towards the rear of the bubble and are dropped off depending on whether they are infront of or behind a critical point similar to the rear horizon for light. This critical point is closer to the ship than the rear horizon, as the massive particles are travelling slower than light and hence find it harder to keep up with the bubble than light does. The critical point is dependant on both the velocity of the ship and the velocity of the massive particles and should be interpreted as the position in the bubble behind which massive particles with such a velocity would not make it to the ship, but would instead be ejected from the rear of the bubble. Additionally, the rear horizon only forms upon the ship reaching the speed of light, however, as the critical point applies to massive particles, it forms (i.e. is inside the bubble) earlier. The velocity of light is constant, so if the ship velocity is below the speed of light then the light eventually reaches the ship. However while in the rear of the bubble, massive particles lose velocity and thus the critical point forms even before the ship reaches the velocity of the massive particles. We can think of the critical velocity defined earlier, as the velocity of massive particles for which the ship velocity generates a critical point at the bubble edge, and thus any massive particles below this velocity attempting to enter the bubble will already be behind the critical point.  

Figure \ref{logistic matter sup rear} shows massive particles with initial velocity 0.8 entering a bubble while the bubble is still travelling subluminally. The massive particles enter the bubble over the period of acceleration, and thus by the time the last several particles have entered the bubble, they are below the critical velocity of the bubble and thus are ejected from the rear of the bubble before the bubble reaches superluminal velocity. 
The remainder of the massive particles with the exception of the first four particles to enter the bubble are one by one overtaken by the critical point and thus are also ejected from the rear of the bubble. 
The later the massive particles are overtaken by the critical point, the greater the reduction in the velocity of the massive particles; this reduction can result in final velocities arbitrarily close to zero. Similarly the later the massive particles ejected due to being below the critical velocity enter the bubble, the smaller the reduction in the final velocity of the massive particles. The combination of these two effects results in massive particles being released from the rear of the bubble with consistently decreasing velocity, which ranges from the initial velocity of the massive particles to arbitrarily close to zero.
Of the massive particles which are not overtaken by the critical point, only the first two to enter the bubble make it to the front horizon before the bubble decelerates below the speed of light and the horizon ceases to exist, and thus only the first two particles to enter the bubble obtain velocities close to unity.

Figure \ref{logistic matter sup front} shows massive particles with velocity 0.1 interacting with a bubble decelerating from a superluminal velocity. This shows how during deceleration massive particles are ejected infront of the bubble at a range of velocities from the initial velocity of the massive particles up to arbitrarily close to the speed of light depending on how long the massive particles have been in the bubble.


\begin{figure}[tbp]
\includegraphics[width=79mm]{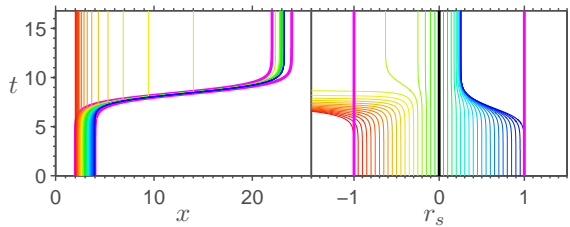}
\caption[Massive particles with zero velocity initially within a bubble on a one way trip]{The left panel is a spacetime diagram of massive particles initially in the bubble with zero velocity interacting with the ship travelling on a subluminal one way journey. The abscissa represents the \(x\) coordinate and the ordinate represents the \(t\) coordinate. The aspect ratio is 1:1 such that light rays travel at \(45^{\circ}\) in flat spacetime. The right panel is the same interaction with respect to the ship, thus the abscissa represents the distance with respect to the ship. The magenta and black lines represent the bubble walls and ship respectively.   \label{logistic matter bubble}}
\end{figure}

Figure \ref{logistic matter bubble} shows the path for massive particles which are spatially stationary and inside the bubble at the origin of the journey. As with light and massive particles discussed earlier, the massive particles from the rear portion of the bubble are dropped off from the bubble with increasing spatial separation. As mentioned earlier, massive particles with zero initial velocity are unaffected by the bubble accelerating or decelerating, and thus upon being dropped off from the bubble they retain a zero velocity. Massive particles in the front portion of the bubble are compressed inwards towards the ship, and massive particles immediately around the ship are largely unaffected. 

\subsection{Superluminal Round Trips}\label{section_round}
\subsubsection{Null Particles}

\begin{figure}[tbp]
\includegraphics[width=79mm]{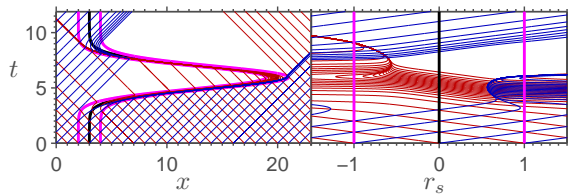}
\caption[Light interacting with a bubble on a round trip]{The left panel is a spacetime diagram of forward and backward travelling light interacting with a ship travelling on a round trip. The abscissa represents the \(x\) coordinate and the ordinate represents the \(t\) coordinate. The aspect ratio is 1:1 such that light rays travel at \(45^{\circ}\) in flat spacetime. The right panel is the same interaction with respect to the ship, thus the abscissa represents the distance with respect to the ship. The blue and red lines are forward and backward travelling light rays respectively, and the magenta and black lines represent the bubble walls and ship respectively. \label{gaussian light}}
\end{figure}

Figure \ref{gaussian light} shows light interacting with a warp bubble on a superluminal round trip. The first part of the trip is very similar to the one way trip shown in Figure \ref{logistic light sup}, with a buildup of forward travelling light at a front horizon and the spreading out and bunching up of backward travelling light rays during the acceleration and deceleration respectively. The second half of the trip is similar to what a one way trip in the opposite direction would look like, with light travelling in the same direction as the ship being captured and light travelling in the opposite direction to the ship passing through the bubble and being bunched up during deceleration at the origin of the trip. A new feature visible is the formation of a front horizon during the second half of the journey with light asymptoting towards it from both sides for much longer than for a normal one way trip, due to the buildup of light in the bubble from the first half of the journey. In both halves of the trip, when the bubble velocity dropped below the speed of light, any light that had been captured was released as a burst.
The path of light is mainly included as reference for the massive particle paths.

\subsubsection{Massive particles}


\begin{figure}[tbp]
\includegraphics[width=79mm]{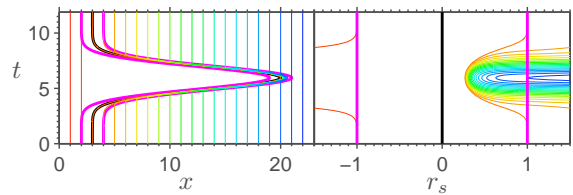}
\caption[Massive particles with zero velocity interacting with a bubble on a round trip]{The left panel is a spacetime diagram of massive particles with zero initial velocity interacting with a ship travelling on a round trip. The abscissa represents the \(x\) coordinate and the ordinate represents the \(t\) coordinate. The aspect ratio is 1:1 such that light rays travel at \(45^{\circ}\) in flat spacetime. The right panel is the same interaction with respect to the ship, thus the abscissa represents the distance with respect to the ship. The magenta and black lines represent the bubble walls and ship respectively. \label{gaussian 0}}
\end{figure}

Figure \ref{gaussian 0} shows the same situation as for light for spatially stationary massive particles. Figure \ref{gaussian 0} shows that the longer the massive particles are inside the bubble, the closer they get to the ship. This is to be expected, as earlier we showed that massive particles with zero initial velocity would eventually pass through a bubble at constant velocity, however this journey is not long enough to allow this to happen. Before the massive particles can pass all the way through the bubble, the bubble begins the return journey. Since the path is symmetric, and the acceleration and deceleration of the bubble has no effect on massive particles with zero initial velocity, as was mentioned earlier, the massive particles leave the bubble at the same spatial positions as they entered the bubble with no change in their velocity. If the journey had been long enough to allow the massive particles to pass through the bubble, any displacement incurred due to passing through the bubble on the first part of the journey would be offset by a corresponding displacement incurred while passing back through the bubble on the second part of the journey. Thus any symmetric path would leave massive particles with zero initial velocity unperturbed. 


\begin{figure}[tbp]
\includegraphics[width=79mm]{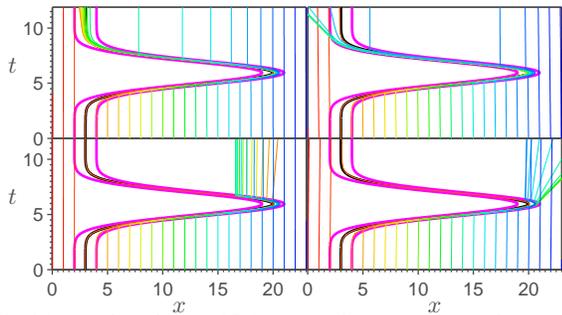}
\vspace{-15mm}
\caption[Massive particles with small negative velocity interacting with a bubble on a round trip]{As in Figure \ref{gaussian 0} for massive particles with nonzero initial velocity. The particles in the upper left panel have velocity -0.001, the upper right -0.01, the lower left 0.001 and the lower right 0.01.\label{gaussian non0}}
\end{figure}

As with the rest of the report, we expect massive particles with nonzero velocity to behave in a similar manner to light, more closely approximating the behaviour of light as the initial velocity of the massive particles approaches that of light. Figure \ref{gaussian non0} shows the same situation for massive particles 
with small and large negative and positive initial velocities. When they leave the bubble again they are spatially displaced in the direction of their initial velocity. 
Thus after the ship has made the round trip, negative velocity particles are shifted towards the origin of the trip in line with how early in the trip the bubble picks them up. This effect could never shift them all to the origin, as it can be thought of as a stretch function with the limiting value of the origin of the trip, no amount of stretching will remove all the negative velocity particles.
Positive velocity particles are captured and asymptote toward the front horizon. The closer the massive particles approach the front horizon, the larger their velocity upon leaving the bubble. This results in a large void between the origin of the journey and part way towards the furthest point away from the origin reached by the bubble, which contains almost none of the massive particles in question. As with the one way trips discussed earlier, a very small amount of massive particles with greatly reduced velocity would be sparsely distributed through this region.
The magnitude of these effects scale with the initial velocity of the particles and the time spent in the bubble, thus particles picked up by the bubble earlier also receive a larger effect.

\section{Conclusions}\label{end}

We have examined the paths of null and massive particles with a range of initial velocities from -\(c\) to \(c\) interacting with an Alcubierre warp bubble travelling at a range of globally subluminal and superluminal velocities on both constant and variable velocity paths.
The key results for null particles match what would be expected of massive test particles travelling at \( \pm 1\). When interacting with a constant velocity warp bubble, particles in the regions N\(^{+}\), N\(^{-}\), and B\(^{+}\) are blueshifted by \(b = 1 - v_{s}v_{p} \) at the ship, but exit the opposite side of the bubble with their original velocity and energy. Particles in the regions P\(^{-}\) and B\(^{-}\) are accelerated and decelerated respectively and ejected out the same side of the bubble that they entered such that there is a bijective map between the regions for the initial and final particle velocity and energy of the particle when interacting with a warp bubble. The blueshift for particle interactions in the P\(^{-}\) and B\(^{-}\)regions are given by \( b =  1 + \frac{v_{s}-v_{p}}{v_{s}}\left( \gamma_{c} - 1 \right) \) and 
\[
b =  \frac{\left(1-\gamma_{c}\right)^{2}+\gamma^{2}_{p}v^{2}_{s}}{\gamma_{c}\gamma^{2}_{p}v^{2}_{s} + \gamma_{p}\sqrt{\left(1-\gamma_{c}\right)^{2} + v^{2}_{s}\left(\gamma^{2}_{p}-\gamma^{2}_{c}\right)}} \nonumber
\]
 respectively.
Particles in the P\(^{+}\) region obtain extremely high blueshifts and become ``time locked" for the duration of their time in the bubble, experiencing very little proper time between entering and eventually leaving the bubble.


When interacting with an accelerating bubble, any particles within the bubble at the time receive a velocity boost that increases or decreases the magnitude of their velocity if the particle is moving towards the front or rear of the bubble respectively. If the bubble is decelerating, the opposite effect is observed. Thus Eulerian matter is unaffected by bubble accelerations/decelerations. Furthermore, if the particle has nonzero initial velocity, then upon leaving the bubble it will still have nonzero velocity. The velocity of null particles cannot be altered by these effects, however the energy of the null particles transforms in the same way as for massive particles. The magnitude of the velocity boosts scales with the magnitude of the bubble acceleration/deceleration. Since these effects occur during the time the particle is inside the bubble, they will have a greater effect on particles that spend more time within the bubble. Hence particles with small negative velocities obtain a larger effect than particles with large negative velocities.
However, the magnitude of the velocity boosts due to bubble acceleration/deceleration are comparatively small to the bubble interaction effects discussed for positive velocity particles, especially when compared to blueshifts found for the P\(^{+}\) region.
The region of space behind a superluminally travelling warp bubble is almost entirely devoid of forward travelling particles, however it contains a sparse distribution of particles with greatly reduced energy. Meanwhile the region of space infront of a ship decelerating from superluminal velocity to subluminal velocity is blasted with a concentrated beam of extremely high energy particles. 

These results suggest that any ship using an Alcubierre warp drive carrying people would need shielding to protect them from potential dangerously blueshifted particles during the journey, and any people at the destination would be gamma ray and high energy particle blasted into oblivion due to the extreme blueshifts for P\(^{+}\) region particles. While in one way journeys particles travelling towards the origin are potentially dangerously blueshifted, their supposed distance from the origin would render them too sparse to be of major concern by the time they reached the origin.   

\begin{acknowledgments}
The authors would like to thank the University of Sydney for their support via the Physics Honours Scholarship.
\end{acknowledgments}


\begin{thebibliography}{18}%
\makeatletter
\providecommand \@ifxundefined [1]{%
 \@ifx{#1\undefined}
}%
\providecommand \@ifnum [1]{%
 \ifnum #1\expandafter \@firstoftwo
 \else \expandafter \@secondoftwo
 \fi
}%
\providecommand \@ifx [1]{%
 \ifx #1\expandafter \@firstoftwo
 \else \expandafter \@secondoftwo
 \fi
}%
\providecommand \natexlab [1]{#1}%
\providecommand \enquote  [1]{``#1''}%
\providecommand \bibnamefont  [1]{#1}%
\providecommand \bibfnamefont [1]{#1}%
\providecommand \citenamefont [1]{#1}%
\providecommand \href@noop [0]{\@secondoftwo}%
\providecommand \href [0]{\begingroup \@sanitize@url \@href}%
\providecommand \@href[1]{\@@startlink{#1}\@@href}%
\providecommand \@@href[1]{\endgroup#1\@@endlink}%
\providecommand \@sanitize@url [0]{\catcode `\\12\catcode `\$12\catcode
  `\&12\catcode `\#12\catcode `\^12\catcode `\_12\catcode `\%12\relax}%
\providecommand \@@startlink[1]{}%
\providecommand \@@endlink[0]{}%
\providecommand \url  [0]{\begingroup\@sanitize@url \@url }%
\providecommand \@url [1]{\endgroup\@href {#1}{\urlprefix }}%
\providecommand \urlprefix  [0]{URL }%
\providecommand \Eprint [0]{\href }%
\providecommand \doibase [0]{http://dx.doi.org/}%
\providecommand \selectlanguage [0]{\@gobble}%
\providecommand \bibinfo  [0]{\@secondoftwo}%
\providecommand \bibfield  [0]{\@secondoftwo}%
\providecommand \translation [1]{[#1]}%
\providecommand \BibitemOpen [0]{}%
\providecommand \bibitemStop [0]{}%
\providecommand \bibitemNoStop [0]{.\EOS\space}%
\providecommand \EOS [0]{\spacefactor3000\relax}%
\providecommand \BibitemShut  [1]{\csname bibitem#1\endcsname}%
\let\auto@bib@innerbib\@empty
\bibitem [{\citenamefont {Alcubierre}(1994)}]{1}%
  \BibitemOpen
  \bibfield  {author} {\bibinfo {author} {\bibfnamefont {M.}~\bibnamefont
  {Alcubierre}},\ }\href {http://stacks.iop.org/0264-9381/11/i=5/a=001}
  {\bibfield  {journal} {\bibinfo  {journal} {Classical and Quantum Gravity}\
  }\textbf {\bibinfo {volume} {11}},\ \bibinfo {pages} {L73} (\bibinfo {year}
  {1994})}\BibitemShut {NoStop}%
\bibitem [{\citenamefont {{Barcel{\'o}}}\ \emph {et~al.}(2010)\citenamefont
  {{Barcel{\'o}}}, \citenamefont {{Finazzi}},\ and\ \citenamefont
  {{Liberati}}}]{17}%
  \BibitemOpen
  \bibfield  {author} {\bibinfo {author} {\bibfnamefont {C.}~\bibnamefont
  {{Barcel{\'o}}}}, \bibinfo {author} {\bibfnamefont {S.}~\bibnamefont
  {{Finazzi}}}, \ and\ \bibinfo {author} {\bibfnamefont {S.}~\bibnamefont
  {{Liberati}}},\ }\href@noop {} {\bibfield  {journal} {\bibinfo  {journal}
  {ArXiv e-prints}\ } (\bibinfo {year} {2010})},\ \Eprint
  {http://arxiv.org/abs/1001.4960} {arXiv:1001.4960 [gr-qc]} \BibitemShut
  {NoStop}%
\bibitem [{\citenamefont {Anderson}\ \emph {et~al.}(2011)\citenamefont
  {Anderson}, \citenamefont {Mackay},\ and\ \citenamefont {Lakhtakia}}]{c33}%
  \BibitemOpen
  \bibfield  {author} {\bibinfo {author} {\bibfnamefont {T.~H.}\ \bibnamefont
  {Anderson}}, \bibinfo {author} {\bibfnamefont {T.~G.}\ \bibnamefont
  {Mackay}}, \ and\ \bibinfo {author} {\bibfnamefont {A.}~\bibnamefont
  {Lakhtakia}},\ }\href {http://stacks.iop.org/2040-8986/13/i=5/a=055107}
  {\bibfield  {journal} {\bibinfo  {journal} {Journal of Optics}\ }\textbf
  {\bibinfo {volume} {13}},\ \bibinfo {pages} {055107} (\bibinfo {year}
  {2011})}\BibitemShut {NoStop}%
\bibitem [{\citenamefont {Ruder}\ \emph {et~al.}(2008)\citenamefont {Ruder},
  \citenamefont {Weiskopf}, \citenamefont {Nollert},\ and\ \citenamefont
  {MÃ¼ller}}]{c27}%
  \BibitemOpen
  \bibfield  {author} {\bibinfo {author} {\bibfnamefont {H.}~\bibnamefont
  {Ruder}}, \bibinfo {author} {\bibfnamefont {D.}~\bibnamefont {Weiskopf}},
  \bibinfo {author} {\bibfnamefont {H.-P.}\ \bibnamefont {Nollert}}, \ and\
  \bibinfo {author} {\bibfnamefont {T.}~\bibnamefont {MÃ¼ller}},\ }\href
  {http://stacks.iop.org/1367-2630/10/i=12/a=125014} {\bibfield  {journal}
  {\bibinfo  {journal} {New Journal of Physics}\ }\textbf {\bibinfo {volume}
  {10}},\ \bibinfo {pages} {125014} (\bibinfo {year} {2008})}\BibitemShut
  {NoStop}%
\bibitem [{\citenamefont {Kobras}\ \emph {et~al.}(2006)\citenamefont {Kobras},
  \citenamefont {Weiskopf}, \citenamefont {Ruder},\ and\ \citenamefont
  {Astrophysik}}]{c21}%
  \BibitemOpen
  \bibfield  {author} {\bibinfo {author} {\bibfnamefont {D.}~\bibnamefont
  {Kobras}}, \bibinfo {author} {\bibfnamefont {D.}~\bibnamefont {Weiskopf}},
  \bibinfo {author} {\bibfnamefont {H.}~\bibnamefont {Ruder}}, \ and\ \bibinfo
  {author} {\bibfnamefont {T.}~\bibnamefont {Astrophysik}},\ }\href@noop {}
  {\bibfield  {journal} {\bibinfo  {journal} {Foundations and Trends® in
  Computer Graphics and Vision}\ }\textbf {\bibinfo {volume} {2}},\ \bibinfo
  {pages} {173} (\bibinfo {year} {2006})}\BibitemShut {NoStop}%
\bibitem [{\citenamefont {Weiskopf}(2000)}]{c22}%
  \BibitemOpen
  \bibfield  {author} {\bibinfo {author} {\bibfnamefont {D.}~\bibnamefont
  {Weiskopf}},\ }in\ \href {\doibase 10.1109/VISUAL.2000.885728} {\emph
  {\bibinfo {booktitle} {Visualization 2000. Proceedings}}}\ (\bibinfo {year}
  {2000})\ pp.\ \bibinfo {pages} {445 --448}\BibitemShut {NoStop}%
\bibitem [{\citenamefont {Clark}\ \emph {et~al.}(1999)\citenamefont {Clark},
  \citenamefont {Hiscock},\ and\ \citenamefont {Larson}}]{8}%
  \BibitemOpen
  \bibfield  {author} {\bibinfo {author} {\bibfnamefont {C.}~\bibnamefont
  {Clark}}, \bibinfo {author} {\bibfnamefont {W.~A.}\ \bibnamefont {Hiscock}},
  \ and\ \bibinfo {author} {\bibfnamefont {S.~L.}\ \bibnamefont {Larson}},\
  }\href {http://www.citebase.org/abstract?id=oai:arXiv.org:gr-qc/9907019}
  {\bibfield  {journal} {\bibinfo  {journal} {Classical and Quantum Gravity}\
  }\textbf {\bibinfo {volume} {16}},\ \bibinfo {pages} {3965} (\bibinfo {year}
  {1999})}\BibitemShut {NoStop}%
\bibitem [{\citenamefont {Pfenning}\ and\ \citenamefont {Ford}(1997)}]{4}%
  \BibitemOpen
  \bibfield  {author} {\bibinfo {author} {\bibfnamefont {M.~J.}\ \bibnamefont
  {Pfenning}}\ and\ \bibinfo {author} {\bibfnamefont {L.~H.}\ \bibnamefont
  {Ford}},\ }\href {doi:10.1088/0264-9381/14/7/011} {\bibfield  {journal}
  {\bibinfo  {journal} {Classical and Quantum Gravity}\ }\textbf {\bibinfo
  {volume} {14}},\ \bibinfo {pages} {1743} (\bibinfo {year}
  {1997})}\BibitemShut {NoStop}%
\bibitem [{\citenamefont {Lobo}\ and\ \citenamefont {Visser}(2004)}]{14}%
  \BibitemOpen
  \bibfield  {author} {\bibinfo {author} {\bibfnamefont {F.~S.~N.}\
  \bibnamefont {Lobo}}\ and\ \bibinfo {author} {\bibfnamefont {M.}~\bibnamefont
  {Visser}},\ }\href {http://stacks.iop.org/0264-9381/21/i=24/a=011} {\bibfield
   {journal} {\bibinfo  {journal} {Classical and Quantum Gravity}\ }\textbf
  {\bibinfo {volume} {21}},\ \bibinfo {pages} {5871} (\bibinfo {year}
  {2004})}\BibitemShut {NoStop}%
\bibitem [{\citenamefont {{Lobo}}(2007)}]{22}%
  \BibitemOpen
  \bibfield  {author} {\bibinfo {author} {\bibfnamefont {F.~S.~N.}\
  \bibnamefont {{Lobo}}},\ }\href@noop {} {\bibfield  {journal} {\bibinfo
  {journal} {ArXiv e-prints}\ } (\bibinfo {year} {2007})},\ \Eprint
  {http://arxiv.org/abs/0710.4474} {arXiv:0710.4474 [gr-qc]} \BibitemShut
  {NoStop}%
\bibitem [{\citenamefont {Davis}(2006)}]{c29}%
  \BibitemOpen
  \bibfield  {author} {\bibinfo {author} {\bibfnamefont {E.~W.}\ \bibnamefont
  {Davis}},\ }in\ \href@noop {} {\emph {\bibinfo {booktitle} {Collection of
  Technical Papers - AIAA/ASME/SAE/ASEE 42nd Joint Propulsion Conference}}},\
  Vol.~\bibinfo {volume} {7}\ (\bibinfo {year} {2006})\ pp.\ \bibinfo {pages}
  {5582--5600}\BibitemShut {NoStop}%
\bibitem [{\citenamefont {{Morris}}\ and\ \citenamefont
  {{Thorne}}(1988)}]{s11}%
  \BibitemOpen
  \bibfield  {author} {\bibinfo {author} {\bibfnamefont {M.~S.}\ \bibnamefont
  {{Morris}}}\ and\ \bibinfo {author} {\bibfnamefont {K.~S.}\ \bibnamefont
  {{Thorne}}},\ }\href {\doibase 10.1119/1.15620} {\bibfield  {journal}
  {\bibinfo  {journal} {American Journal of Physics}\ }\textbf {\bibinfo
  {volume} {56}},\ \bibinfo {pages} {395} (\bibinfo {year} {1988})}\BibitemShut
  {NoStop}%
\bibitem [{\citenamefont {{Coule}}(1998)}]{5}%
  \BibitemOpen
  \bibfield  {author} {\bibinfo {author} {\bibfnamefont {D.~H.}\ \bibnamefont
  {{Coule}}},\ }\href {\doibase 10.1088/0264-9381/15/8/026} {\bibfield
  {journal} {\bibinfo  {journal} {Classical and Quantum Gravity}\ }\textbf
  {\bibinfo {volume} {15}},\ \bibinfo {pages} {2523} (\bibinfo {year}
  {1998})}\BibitemShut {NoStop}%
\bibitem [{\citenamefont {Olum}(1998)}]{s8}%
  \BibitemOpen
  \bibfield  {author} {\bibinfo {author} {\bibfnamefont {K.~D.}\ \bibnamefont
  {Olum}},\ }\href {\doibase 10.1103/PhysRevLett.81.3567} {\bibfield  {journal}
  {\bibinfo  {journal} {Phys. Rev. Lett.}\ }\textbf {\bibinfo {volume} {81}},\
  \bibinfo {pages} {3567} (\bibinfo {year} {1998})}\BibitemShut {NoStop}%
\bibitem [{\citenamefont {{Visser}}\ \emph {et~al.}(2000)\citenamefont
  {{Visser}}, \citenamefont {{Bassett}},\ and\ \citenamefont
  {{Liberati}}}]{s9}%
  \BibitemOpen
  \bibfield  {author} {\bibinfo {author} {\bibfnamefont {M.}~\bibnamefont
  {{Visser}}}, \bibinfo {author} {\bibfnamefont {B.~A.}\ \bibnamefont
  {{Bassett}}}, \ and\ \bibinfo {author} {\bibfnamefont {S.}~\bibnamefont
  {{Liberati}}},\ }\href {\doibase 10.1016/S0920-5632(00)00782-9} {\bibfield
  {journal} {\bibinfo  {journal} {Nuclear Physics B Proceedings Supplements}\
  }\textbf {\bibinfo {volume} {88}},\ \bibinfo {pages} {267} (\bibinfo {year}
  {2000})},\ \Eprint {http://arxiv.org/abs/arXiv:gr-qc/9810026}
  {arXiv:gr-qc/9810026} \BibitemShut {NoStop}%
\bibitem [{\citenamefont {{Low}}(1999)}]{s10}%
  \BibitemOpen
  \bibfield  {author} {\bibinfo {author} {\bibfnamefont {R.~J.}\ \bibnamefont
  {{Low}}},\ }\href {\doibase 10.1088/0264-9381/16/2/016} {\bibfield  {journal}
  {\bibinfo  {journal} {Classical and Quantum Gravity}\ }\textbf {\bibinfo
  {volume} {16}},\ \bibinfo {pages} {543} (\bibinfo {year} {1999})},\ \Eprint
  {http://arxiv.org/abs/arXiv:gr-qc/9812067} {arXiv:gr-qc/9812067} \BibitemShut
  {NoStop}%
\bibitem [{Note1()}]{Note1}%
  \BibitemOpen
  \bibinfo {note} {These equations were numerically integrated to find \(t\),
  \(x\), \(u^{t}\) and \(u^{x}\) of the particles using the inbuilt Ordinary
  Differential Equation (ODE) solver ode23s in Matlab. Only one pair of these
  variables (\(t\) and \(u^{t}\), or \(x\) and \(u^{x}\)) needed to be solved
  for, however both were found and the normalisation of the 4 velocity was used
  as a check on the accuracy of the integration. In cases where \(u^{t}\) and
  \(u^{x}\) diverge, the equations were numerically integrated to find x and
  \({\begingroup dx\endgroup \over dt}\) using ode45.}\BibitemShut {Stop}%
\bibitem [{\citenamefont {Narlikar}(1994)}]{energy}%
  \BibitemOpen
  \bibfield  {author} {\bibinfo {author} {\bibfnamefont {J.~V.}\ \bibnamefont
  {Narlikar}},\ }\href {\doibase DOI:10.1119/1.17679} {\bibfield  {journal}
  {\bibinfo  {journal} {American Journal of Physics}\ }\textbf {\bibinfo
  {volume} {62}},\ \bibinfo {pages} {903} (\bibinfo {year} {1994})}\BibitemShut
  {NoStop}%
\end{thebibliography}

%

\end{document}